\documentclass[
superscriptaddress,
reprint,
 amsmath,amssymb,
 aps,
 prx,
]{revtex4-1}

\usepackage{graphicx}
\usepackage{dcolumn}
\usepackage{bm}
\usepackage{amsmath}
\usepackage{amsfonts}
\usepackage{amssymb}
\usepackage{dsfont}
\usepackage{color}
\usepackage{xcolor}
\usepackage{siunitx}
\usepackage[colorlinks=true,linkcolor=blue]{hyperref}
\usepackage{braket}
\usepackage{mathtools}
\usepackage[normalem]{ulem}
\usepackage{scalerel}
\usepackage{ bbold }
\usepackage{inputenc}
\usepackage[T1]{fontenc}

\definecolor{mygreen}{rgb}{0,0.5,0}
\definecolor{myblue}{rgb}{0,0,0.75}
\definecolor{mymagenta}{cmyk}{0,1,0,0.12}
\definecolor{mygray}{rgb}{0.5,0.5,0.5}

\newcommand{\overbar}[1]{\mkern 1.5mu\overline{\mkern-1.5mu#1\mkern-1.5mu}\mkern 1.5mu} 

\graphicspath{{./figures/}}

\begin{document}

\title{Universal quantum computation and quantum error correction \\ with ultracold atomic mixtures}

\author{Valentin Kasper}
\affiliation{ ICFO-Institut  de  Ciencies  Fotoniques,  The  Barcelona  Institute  of  Science  and  Technology,Av.   Carl  Friedrich  Gauss  3,  08860  Barcelona,  Spain}
\email{valentin.kasper@icfo.eu}

\author{Daniel González-Cuadra}
\affiliation{ ICFO-Institut  de  Ciencies  Fotoniques,  The  Barcelona  Institute  of  Science  and  Technology,Av.   Carl  Friedrich  Gauss  3,  08860  Barcelona,  Spain}

\author{Apoorva Hegde}
\affiliation{Universit\"at  Heidelberg,  Kirchhoff-Institut  f\"ur  Physik, Im  Neuenheimer  Feld  227,  69120  Heidelberg,  Germany }

\author{Andy Xia}
\affiliation{Universit\"at  Heidelberg,  Kirchhoff-Institut  f\"ur  Physik, Im  Neuenheimer  Feld  227,  69120  Heidelberg,  Germany }

\author{Alexandre Dauphin}
\affiliation{ ICFO-Institut  de  Ciencies  Fotoniques,  The  Barcelona  Institute  of  Science  and  Technology,Av.   Carl  Friedrich  Gauss  3,  08860  Barcelona,  Spain}

\author{Felix Huber}
\affiliation{ ICFO-Institut  de  Ciencies  Fotoniques,  The  Barcelona  Institute  of  Science  and  Technology,Av.   Carl  Friedrich  Gauss  3,  08860  Barcelona,  Spain}

\author{Eberhard Tiemann}
\affiliation{Institut für Quantenoptik, Leibniz Universität Hannover, 30167 Hannover, Germany}

\author{Maciej Lewenstein}
\affiliation{ ICFO-Institut  de  Ciencies  Fotoniques,  The  Barcelona  Institute  of  Science  and  Technology,Av.   Carl  Friedrich  Gauss  3,  08860  Barcelona,  Spain}
\affiliation{ICREA, Pg. Lluís Companys 23, 08010 Barcelona, Spain}

\author{Fred Jendrzejewski}
\affiliation{Universit\"at  Heidelberg,  Kirchhoff-Institut  f\"ur  Physik, Im  Neuenheimer  Feld  227,  69120  Heidelberg,  Germany }

\author{Philipp Hauke}
\affiliation{INO-CNR  BEC  Center  and  Department  of  Physics,University  of  Trento,  Via  Sommarive  14,  I-38123  Trento,  Italy}

\date{\today}

\begin{abstract}
Quantum information platforms made great progress in the control of many-body entanglement and 
the implementation of quantum error correction, but it remains a challenge to realize both in the same setup. Here, we propose a mixture of two ultracold atomic species as a platform for universal quantum computation with long-range entangling gates, while providing a natural candidate for quantum error-correction. 
In this proposed setup, one atomic species realizes localized collective spins of tunable length, which form the fundamental unit of information. The second atomic species yields phononic excitations, which are used to entangle collective spins. 
Finally, we discuss a finite-dimensional version of the Gottesman-Kitaev-Preskill code to protect quantum information encoded in the collective spins, opening up the possibility to universal fault-tolerant quantum computation in ultracold atom systems.

\end{abstract}

\maketitle
\begin{figure}[t!]
    \centering
    \includegraphics[width=\columnwidth]{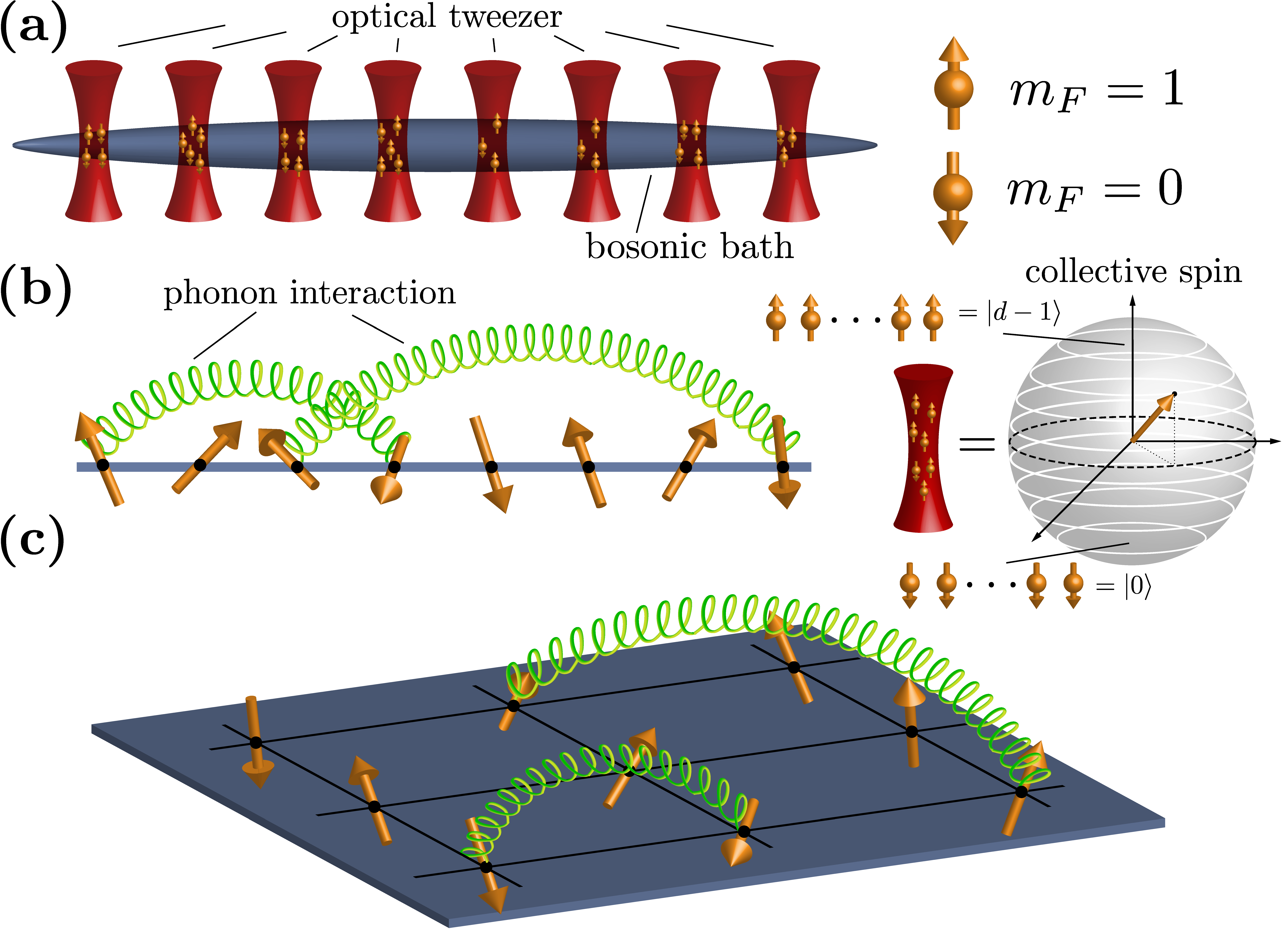}
    \caption{\textbf{Quantum computation with ultracold atomic mixtures.} 
    \textbf{(a)} Proposed experimental platform for the case of one dimension. Bosonic atoms A (yellow) are trapped via optical potentials (red tweezers) and immersed in a one dimensional quasi-condensate.
    formed by bosonic atoms B  (blue).
    \textbf{(b,c)} One/two-dimensional spin-phonon model. Two hyperfine levels of A atoms form  collective spins (yellow arrows), which interact via phonon excitations of B atoms (green lines). 
    }

    \label{fig:Overview}
\end{figure}

Quantum information processing is expected to show fundamental advantages over classical approaches to computation, simulation, and communication~\cite{Acin2018}. 
To exploit these advantages, both the efficient creation of entanglement and the correction of errors are essential \cite{Horodecki2009, Terhal2015}. 
There has been spectacular advancements in the field of synthetic quantum devices, which allows for controlled many-body entanglement~\cite{Bernien2017, Zhang2017, Arute2019} or the implementation of quantum error correction~\cite{Nigg2014, Fluhmann2019, Campagne-Ibarcq2019, Andersen2020}. 
However, it is still a challenge to design a quantum computation platform
that efficiently realizes both features simultaneously. 
Here, we propose to use a mixture of two ultracold atomic species to implement a universal quantum computer, permitting long-range entangling gates as well as quantum error correction (see Fig.~\ref{fig:Overview}). One atomic species forms localized collective spins, which can be fully controlled and represent the basic units of information. The second species is used for the generation of pairwise entanglement of the collective spins, resulting in a universal gate set. Moreover, we illustrate how to encode a logical qubit in the collective spin, how to detect and correct errors, and present a universal gate set on the logical qubits. 
Altogether, this opens a new possibility for quantum computation with error correction in ultracold atomic systems, presenting a significant step towards fault-tolerant quantum computation. 

The proposal of this work complements existing platforms such as nuclear magnetic resonance~\cite{Vandersypen2001}, nitrogen-vacancy centers in diamonds~\cite{Doherty2013}, photonics~\cite{Knill2001}, silicon based qubits~\cite{Watson2018}, superconducting qubits~\cite{Krantz2019}, and trapped ions~\cite{Bruzewicz2019}. 
One key challenge {in many of these platforms} is to reduce
the technical overhead when synthesizing long-range gates out of short range gates, as it happens, e.g., in superconducting qubit experiments. 
While trapped ion systems provide efficient long-range entangling gates, they face technical challenges in establishing full control beyond  one hundred ions~\cite{Nigmatullin2016, Pagano2018}. 
Furthermore, many implementations of quantum error correction demand 
a considerable number of physical qubits to store one logical qubit, and additional 
control qubits for the encoding, decoding, and correcting processes~\cite{Campbell2017}.
Whereas the challenges of efficient long-range entanglement and quantum error correction appear daunting, ultracold atomic mixtures may provide a solution for both.  

Ultracold atoms have become a major quantum simulation platform to solve problems in the field of condensed-matter~\cite{Bloch2008, Cirac2012,Lewenstein2012} and high-energy physics~\cite{Gorg2019, Schweizer2019, Mil2019, Yang2020, Banuls2020}. 
Yet, even though universal quantum information processing with ultracold atom systems was investigated conceptually, the experimental implementations remain elusive~\cite{Jaksch2004, BrickmanSoderberg2009, Saffman2010}. The realization of gates with high-fidelity and the ability to apply {sequences of} several gates is particularly challenging. 
However, recent progress in 
{systems of Rydberg atoms trapped in optical tweezers}
has made it possible to realize 
large quantum many-body systems~\cite{Omran2019} with fast, high-fidelity entangling gates~\cite{Madjarov2020}. 
Simultaneously, multi-component Bose--Einstein condensates were used to form large collective  
spins~\cite{Strobel2014, Lucke2014, Bookjans2011, Pezze2018, Stamper-Kurn2013} and spatially distribute entanglement via
expansion~\cite{Kunkel2018}. 
Ultracold atomic mixtures allow for another entanglement mechanism 
based on phonon induced interactions. 
The effect of phonons on a single atomic species has already been experimentally studied 
in the context of polaron physics~\cite{Klein2005, Scelle2013a, Rentrop2016}. 


We first introduce the fundamental unit of information of our platform: collective spins of controllable length. 
They are realized by condensing bosonic atoms into a single spatial mode of an optical tweezer. 
The remaining degrees of freedom are two internal states of the atoms, 
which can be described using a collective spin.  The length of the collective spin is controlled by
the number of trapped atoms enabling access to the qubit, qudit, and continuous variable regime. The collective occupation of the internal degrees of freedom 
constitute the computational basis. Tuning the length of the collective spin, 
quantum information processing can be done with qubits, qudits or continuous variables. The  atoms within a tweezer can be reliably prepared in a fully polarized state,
which then acts as the initial state of the computation. 
A single collective spin can be fully controlled via linear operations, e.g., realized by a microwave field, and non-linear operations such as the interaction between the atoms in different hyperfine states. 
In order to create entanglement between the collective spins, we employ a second atomic species, 
which forms a bosonic bath with phononic excitations. The contact interaction between the two atomic species gives rise to the exchange of phononic excitations between the collective spins. The resulting long-range interaction permits an efficient generation of entanglement between distant spins. The operations on a single collective 
spin in combination with their pairwise entanglement forms a universal gate set. 
The expected gate speed is to be much faster then the decoherence time in the platform suggested here. 
Altogether, this platform fulfills DiVinvenzo's criteria for quantum computation~\cite{DiVincenzo2000}.

Moreover, we illustrate a scheme for quantum error correction based on the stabilizer formalism.
The method encodes a logical qubit  into superpositions of states within the higher dimensional Hilbert space of the collective spin. We further explain how to prepare, detect and correct specific errors by using additional control qubits. Finally, we present a universal gate set for the logical qubits.

This work is organized as follows. In Sec.~\ref{sec:EffectiveSpins}, we explain how one atomic species forms a collective spin and how to generate arbitrary unitaries on such a single object. In Sec.~\ref{sec:Entanglement}, we discuss the interaction between the collective spins and a phonon bath, and how this can be used to entangle two collective spins. This in turn can then be employed to achieve universal quantum computation. 
In Sec.~\ref{sec:StatePreparation}, we discuss the state preparation and readout of the collective spins. The  experimental characteristics of this quantum computation platform are discussed in Sec.~\ref{sec:Experimental}. Finally, in Sec.~\ref{sec:QEC} we discuss how to implement quantum error correction in this spin-phonon platform.

\section{Collective Spins\label{sec:EffectiveSpins}}
In this section, we discuss the fundamental unit of information of our proposal: a collective spin with controllable length. 
In order to realize such a collective spin we consider bosonic atoms of type A, which are 
localized at the local minima  $\mathbf{y}$ of the optical potential, e.g., arrays of optical tweezer or an optical lattice (see Fig. \ref{fig:Overview}a).
The annihilation (creation) operators of the atoms on site $\mathbf{y}$ are $a_m(\mathbf{y})$ and $a^{\dagger}_m(\mathbf{y})$, where $m$ indicates the two internal states (0, 1) of the atom. 
The local confining potential is chosen such that there is no hopping. Hence, the atom number per site $N_A(\mathbf{y}) =a_{1}^{\dagger } (\mathbf{y}) a_{1} (\mathbf{y}) + a_{0}^{\dagger } (\mathbf{y})  a_{0}(\mathbf{y})$ is conserved~\footnote{
If not stated
otherwise we will assume the same number of atoms at all site $\mathbf{y}$, whereas 
different atom number per sites are in principle possible} and the atoms form a collective spin via the Schwinger representation:
\begin{subequations} \label{eq:SchwingerRepresentation}
\begin{align}
    L_{z}(\mathbf{y}) &=\frac{1}{2} [ a_{1}^{\dagger} (\mathbf{y}) a_{1} (\mathbf{y}) - a_{0}^{\dagger }(\mathbf{y}) a_0(\mathbf{y}) ] \, ,\\
    L_{+}(\mathbf{y}) &= a_{1}^{\dagger} (\mathbf{y}) a_{0} (\mathbf{y}) \, , \\
    L_{-}(\mathbf{y}) &= a_{0}^{\dagger} (\mathbf{y}) a_{1} (\mathbf{y}) \, ,
\end{align}
\end{subequations}
with $L_{x}(\mathbf{y})=[L_{+}(\mathbf{y})+L_{-}(\mathbf{y})]/2$ and $L_{y}(\mathbf{y})=(-i)[ L_{+}(\mathbf{y})-L_{-}(\mathbf{y})]/2$.
The conservation of atom number per site determines 
also the magnitude of the angular momentum $\ell = N_A/2$. 
The eigenstates of the $L_{z}$ operator are denoted by $\ket{m_{\ell}}$ with $m_{\ell}$ being an 
integer or half-integer ranging from $m_{\ell}=-\ell, \ldots, \ell$. By defining the
computational basis $\ket{j} \equiv \ket{-\ell + j}$ 
with $j = 0, \ldots, N_A $ a collective spin can be interpreted
as a qudit  with dimension $d = N_A + 1$. Moreover, we can access the qubit ($N_A = 1$), qudit ($N_A>1$), and continuous variable ($N_A\rightarrow \infty$) regime by tuning the atom number per site.
In particular, the  Hilbert space dimension of a single collective spins scales {linearly} with the numbers of atoms. 

In order to realize unitary operations acting on a single collective spin, we consider the
time evolution generated by the Hamiltonian 
\begin{align}
    {{H}_{A}}\!=\!\mathop{\sum }_{\mathbf{y}}\left[{\chi}(\mathbf{y}) L^2_z(\mathbf{y})+\Delta(\mathbf{y})L_{z}({\mathbf{y}}) +   \Omega(\mathbf{y}) L_x(\mathbf{y})\right]\!. \label{eq:HamiltonianA}
\end{align}
As detailed in App.~\ref{sec:SingleHamiltonian}, this Hamiltonian can be implemented by a two component Bose gas localized in optical tweezers, where the first term corresponds to the interaction between the atoms, 
the second is due to the presence of the magnetic field, 
and the third term to a Rabi coupling between the hyperfine states.

The Hamiltonian in Eq.~\eqref{eq:HamiltonianA} can generate all unitary operations 
on a single collective spin, if one assumes that the couplings 
${\chi}(\mathbf{y})$, $\Delta(\mathbf{y})$, and $\Omega(\mathbf{y})$ can be switched on and off independently.
For the qubit case ($d=N_A+1=2$), the operators $L_x$ and $L_z$ together with the identity $\mathbb{1}$ suffice to obtain all $U(2)$ transformations~\cite{Nielsen2013}.  
For qudits ($d=N_A+1>2$) the additional non-linear operation $(L_z)^2$ is required to approximately synthesize any unitary $U(d)$ via Trotterization. 
Namely, using $e^{-i \hat{A} \delta t} e^{i \hat{B} \delta t} e^{i \hat{A} \delta t} e^{-i \hat{B} \delta t} = e^{[\hat{A}, \hat{B}] \delta t^{2}}+O\left(\delta t^{3}\right)$, we can engineer all possible commutators of $L_x$, $L_z$, and $L_z^2$ and higher order commutators. These commutators span the $d$-dimensional Hermitian matrices with the basis $\{\mathcal{M}_i\}_{i=1}^{d^2}$, which in turn allows to synthesize all unitary matrices.
In App.~\ref{sec:EnvelopingAlgebra} we give an explicit construction of the enveloping algebra of $L_x$, $L_z$, and $L_z^2$ for $d = 3$, which allows one to generate each element of $U(3)$. 
As detailed in Ref.~\cite{Giorda2003}, the construction given in the appendix can be generalized to $d>3$. 
In this way, the operators $L_x$, $L_z$, and $L_z^2$ provide universal control over each single collective spin. 
In the next section, we consider the entanglement of several collective spins to achieve exponential growth of the Hilbert space.

\section{Unitary operations on two spins \label{sec:Entanglement}}
The entanglement of different collective spins can be achieved by the exchange of delocalized 
phonons. The phonons are the Bogoliubov excitations of a weakly interacting condensate of atomic species B in $n$ dimensions, 
where we consider $n = 1,2$ (see Fig.~\ref{fig:Overview}). 
The purely phononic part of the Hamiltonian is given by 
\begin{align}
H_B =& \sum_{\mathbf{k}} \hbar \omega_{\mathbf{k}} b^{\dagger}_{\mathbf{k}} b_{\mathbf{k}} \,, \label{eq:PhononHamiltonian}
\end{align}
where $b_{\mathbf{k}}$ is the annihilation operator of a phonon mode at wave number $\mathbf{k}$ with components $k_i = \frac{\pi}{L}m_i$ and $m_i$ a positive integer.
The phonon dispersion at low momenta  is linear $\omega_{\mathbf{k}}\approx c |\mathbf{k}|$,
where $c= \sqrt{\tilde{g}_B n_B/M_B}$ is the speed of sound determined by the 
interaction strength $\tilde{g}_B$ of the B atoms with mass $M_B$ and the density $n_B$. See App.~\ref{Sec:Phonon} for more details.

By immersing the collective spins into the phonon bath,  
we induce interactions between the phonons and the spins 
rooted in the contact interactions of the two atomic species A and B. 
The spin-phonon interaction can be described by the Hamiltonian 
\begin{align}
{H}_{AB} =&  \sum_{\mathbf{y},\mathbf{k}} \left[\bar{g}_{\mathbf{k}}(\mathbf{y}) +  \delta g_\mathbf{k}(\mathbf{y}) L_z(\mathbf{y}) \right] (b_{\mathbf{k}} + \text{H.c.}) \, , \label{eq:SpinPhononHamiltonian}
\end{align}
with the explicit forms of the coupling constants $\bar{g}_{\mathbf{k}}(\mathbf{y})$ and $g_\mathbf{k}(\mathbf{y})$
as well as a detailed derivation given in App.~\ref{app:SpinPhonon}. 
This interaction is similar to the phonon-ion ineractions in trapped ion systems~\cite{Bruzewicz2019} or photon-atom interactions~\cite{Chang2018}.
The first term of Eq.~\eqref{eq:SpinPhononHamiltonian} leads to a constant polaronic shift ~\cite{Grusdt2015, Rentrop2016},
which can be absorbed by redefining the phonon operators.
Consequently, we focus here on the second term, which can be used to generate entanglement between the spins.

Since $L_z(\mathbf{y})$ is a conserved quantity for $\Omega(\mathbf{y}) =0$ it is 
possible to decouple the phonons and spins in Eq.~\eqref{eq:SpinPhononHamiltonian}\, by 
shifting the phonon operators (see App.~\ref{app:SpinPhonon}). 
Eliminating the phonons leads to an effective long-range spin-spin interaction
\begin{align} \label{Eq:ZZInteraction}
H_{I} =&   \sum_{\mathbf{x}, \mathbf{y}} {g}(\mathbf{x},\mathbf{y}) L_z(\mathbf{x}) L_z(\mathbf{y})\, , 
\end{align}
where we introduced the coupling ${g}(\mathbf{x},\mathbf{y})$ between the spins  after a proper
redefinition of $\chi(\mathbf{y})$ and $\Delta(\mathbf{y})$. 
\begin{figure}
    \centering
    \includegraphics[width=\columnwidth]{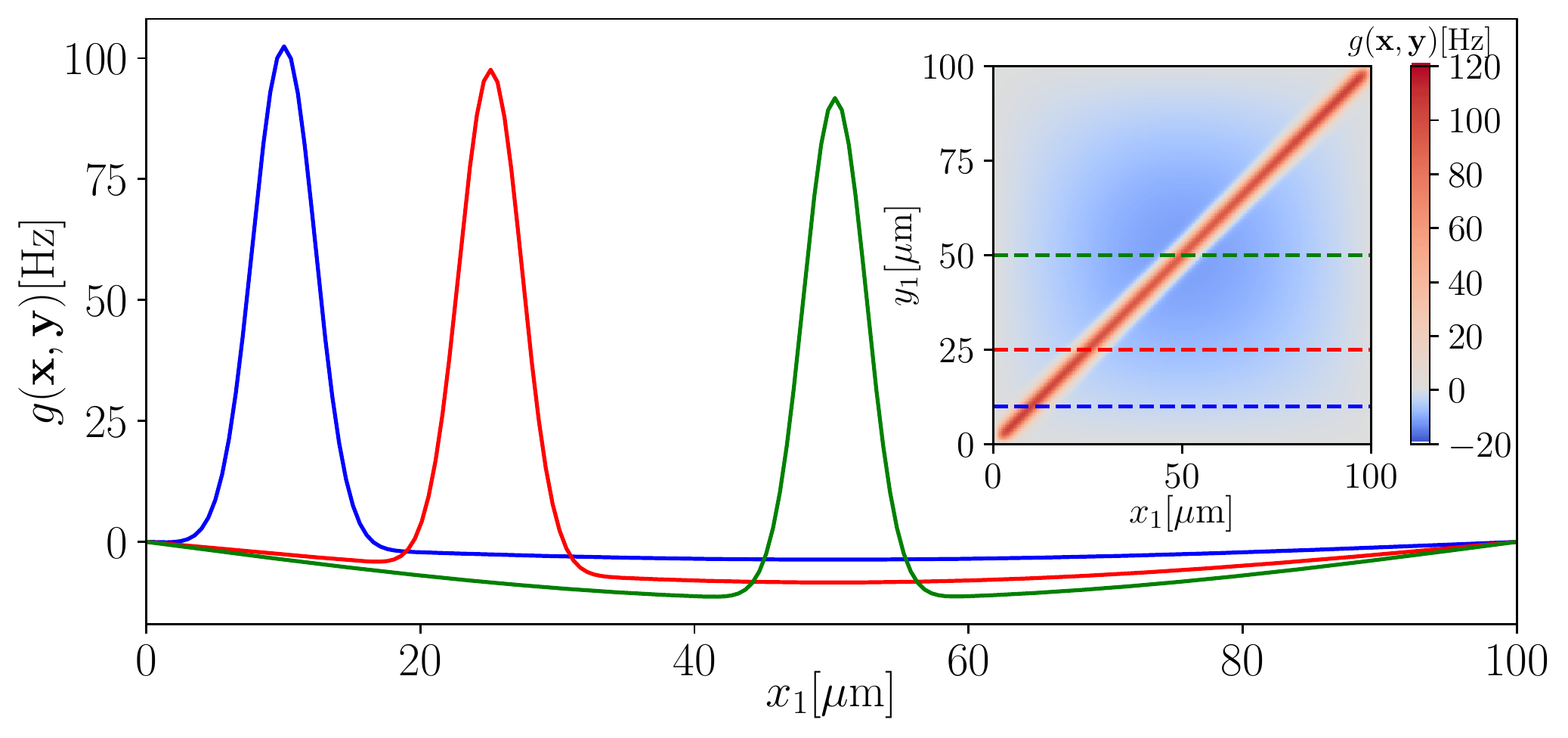}
    \caption{\textbf{Spin-spin interaction in one dimension:} 
    One collective spin is located at positions $y_i$, 
    while the other collective spin is located at position $y_j$. 
    The interaction between spins at $y_i$ and $y_j$ decays towards the boundary
    of the box and for specific positions changes its sign. 
    For our proposal, the collective spins do not overlap, because of the {small} spatial extent of the tweezer (see App.~\ref{sec:SingleHamiltonian} for details). 
    Hence, we focus on the region with $|y_i-y_j| \gg \sigma_A$, where $\sigma_A$ is the length scale of
    the confining harmonic oscillator.
    \textbf{(Inset)} The phonon-mediated interaction decays when moving away from the collective spin
    and becomes zero at the boundary due to the box potential 
    (Dirichlet boundary conditions). 
    The sign of the interaction can be tuned by changing the relative position of the
    collective spins. The colored dashed lines correspond to the cuts in the outer 
    figure.
    \label{fig:SpinPhonoModel1D}    }
\end{figure}

Explicitly, the coupling between the spins is given by
\begin{align}
    \! g(\mathbf{x},\mathbf{y}) \!=\! g \sum_{\mathbf{k}}\!\frac{(u_{\mathbf{k}} + v_{\mathbf{k}})^2}{\hbar \omega_{\mathbf{k}}}  e^{-\frac{1}{4} |\mathbf{k}|^2 l_{\!\scriptscriptstyle{A}}^2}\! \prod_{i=1}^n \sin(k_i x_i) \sin(k_i y_i)
    \label{eq:effectiveInteraction} \,.
\end{align}
The overall prefactor
\begin{align}
g &= n_B  \left( 2/L\right)^{n} (\tilde{g}^1_{AB} - \tilde{g}^0_{AB})^2 \, 
\end{align}
is determined by the inter-species interaction $\tilde{g}^0_{AB}$ and $\tilde{g}^1_{AB}$, where
the $\tilde{}$ indicates the renormalization according to the optical potential. 
Further, we introduce the Bogoliubov amplitudes $u_{\mathbf{k}}$ and $v_{\mathbf{k}}$, which
are determined by performing the Bogoliubov approximation in a box potential of length $L$~\cite{Lopes2017,Rauer2018}. We further approximate the Bogoliubov mode functions by $\sin(k_ix_i)$.
The length scale $\sigma_A$ is the harmonic oscillator length of the tweezer confining the $A$ atoms, 
which provides a cutoff for the momentum sum in Eq.~\eqref{eq:effectiveInteraction}. For more details we refer to App.~\ref{app:SpinPhonon}.

\begin{figure}
    \centering
    \includegraphics[width=\columnwidth]{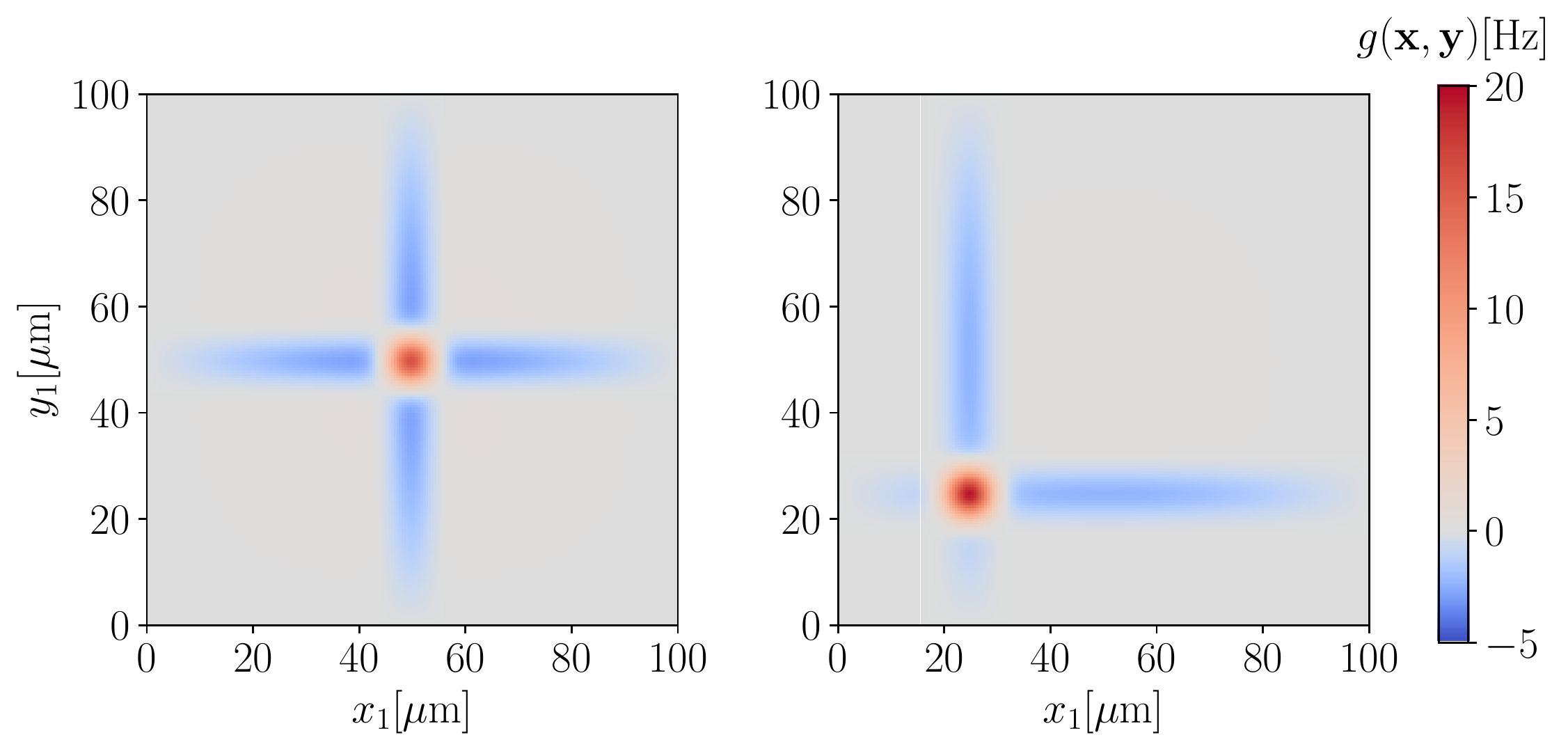}
    \caption{\textbf{Spin-spin interaction in two dimension:} The collective spin is immersed in
    a two dimensional bosonic bath and located at the center of the box \textbf{(a)} or at the lower left corner \textbf{(b)}. Similarly to the one-dimensional
    case the interaction is long-ranged, decays when moving away from the impurities, and vanishes at the boundary of the system. The interaction can be tuned by choosing the 
    relative position of the two collective spins. }
    \label{fig:SpinPhonoModel2D}
\end{figure}

Performing the momentum sum numerically, the resulting interaction between two spins for one and two dimensions is illustrated in Fig.~\ref{fig:SpinPhonoModel1D} and Fig.~\ref{fig:SpinPhonoModel2D}. 
Positioning the spins appropriately within the bath allows one to tune the strength and sign of the spin-spin interaction. The effective interaction scales as $\sqrt{n_B}$ independent of the dimension, as long as only the linear
part of the dispersion $\omega_{\mathbf{k}}$ contributes to the sum of Eq.~\eqref{eq:effectiveInteraction}.

In order to entangle two specific spins, the ability to deliberately switch on and off the interactions $g(\mathbf{x}, \mathbf{y})$ is necessary. 
A first possibility is to physically move the optical tweezers such that there is no overlap of the collective spins with the phonons. This approach is similar in spirit to the shuttling approach in trapped ions~\cite{Kaushal2020}. 

The second approach, is similar to the optical shelving used in trapped ions~\cite{Riebe2004}. The interaction between the collective spins is proportional to the scattering length difference, $g(\mathbf{x},\mathbf{y})\propto \left( \tilde{g}^1_{AB} - \tilde{g}^0_{AB}\right)^2$. 
The first term describes the interaction strength between A atoms in $m_F=1$ and B atoms in $m_F=0$, while the second term describes the interaction between atoms of both species being in $m_F=0$. To shelve spins, we can coherently transfer the $m_F=0$ component of A atoms into the $m_F=-1$ state, while the atoms in the $m_F=1$ component are left unaltered. The resulting interaction is strictly zero since $g(\mathbf{x},\mathbf{y})\propto \left( \tilde{g}^{-1}_{AB} - \tilde{g}^1_{AB}\right)^2 = 0$. 

Tuning the interaction time between the spin plus two single qudit gates and a phase provide a controlled-Z gate  
\footnote{The definition of a d-dimensional $\operatorname{CZ}$ gate is given by 
$\operatorname{CZ}\ket{p}\ket{q} = \omega^{pq} \ket{p}\ket{q}$ }
between two collective spins
\begin{align} \label{eq:CZgate}
    \operatorname{CZ}(\mathbf{x},\mathbf{y})\!=\! e^{i \frac{2\pi}{d} L_z(\mathbf{x})L_z(\mathbf{y})} e^{i \frac{2\pi \ell}{d}L_z(\mathbf{x})} e^{i \frac{2\pi \ell}{d}L_z(\mathbf{y})} e^{i \frac{2\pi \ell^2}{d}} ,
\end{align}
which completes the universal gate set in the multi-spin system~\cite{LLoyd1995, Bartlett2002}.
The quality of the local spin addressing and the magnetic field stability will determine the fidelity of the gates.

\section{State preparation and detection \label{sec:StatePreparation}}
A universal quantum computer requires a reliable state-preparation and readout. Using the external magnetic field one can prepare all A atoms in one hyperfine component, which corresponds to a fully polarized collective spin along the quantization axis $\ket{\psi} = \ket{d} \otimes \ldots \ket{d}$. In case the particle number per site is probabilistic, one can perform post-selection to fix $N_A$~\cite{Endres2016, Barredo2016, Wang2020}. If one uses instead an optical lattice to create the collective spins, one can deterministically control the atom number by preparing first a Mott-insulating state, which nowadays can be prepared with almost unit filling~\cite{Yang2020} and a subsequent merging of a fixed number of wells. For the preparation as well as for the detection one has to ensure single counting statistics per site, which can be achieved through fluorescence imaging~\cite{Strobel2014}.

When working in the large collective spin regime, one can efficiently prepare condensates through evaporative cooling in the optical dipole trap~\cite{Muessel2014}. This results in an atomic cloud of a few hundred to thousand atoms of low entropy in each dipole trap. In case the A atoms are not in the motional ground state, one can perform Raman sideband cooling~\cite{Kaufman2012}. For larger collective spins, one can use homodyne detection to map out large collective spins~\cite{Gross2011}.

\section{Experimental characteristics \label{sec:Experimental}} 
Having established all the necessary ingredients for a quantum information platform, we discuss details of a possible experimental realization as well as main sources of errors. The experiments can rely on standard experimental tools employed for cold atoms~\cite{Bloch2008}. For concreteness, we focus on a mixture of \textsuperscript{39}K (species A) forming the collective spins and \textsuperscript{23}Na (species B) forming the phonon bath. 

The collective spins can be realized through the trapping of a few atoms in optical tweezers distanced by a few micrometers and with motional ground state extension $\sigma_A\approx \SI{100}{\nano \meter}$.  The typical single qubit gates can then be performed with a Rabi frequency of up to a few \si{\kilo \Hz} with above $99\%$ fidelity~\cite{Stamper-Kurn}. Readout reliability through fluorescence imaging differ between schemes but can reach $>99\%$ fidelity~\cite{Boll2016a, Bergschneider2019, Covey2019}. 

Assuming a box potential for species B~\cite{Lopes2017} in a tube or a slab geometry with linear dimension $L \approx \SI{100}{\micro \meter}$ we expect approximately 50 collective spins in a one-dimensional geometry or 2500 collective spins in a two dimensional geometry. In both cases, all-to-all coupling can be achieved through phonons of a (quasi-)condensate of approximately $N_B \approx 300 \times 10^3$ atoms for the 1D bath and $N_B \approx 3 \times 10^6$ for the 2D bath, with a transverse confinement of $\omega_\perp \approx 2\pi \times \SI{440}{\Hz}$, such that the condensate has a chemical potential of $\mu_B/h \approx \SI{7.7}{\kHz}$ in 1D and $\mu_B/h \approx \SI{1.9}{\kHz}$ in 2D~\cite{Lewenstein2012}. This implies typical lifetimes for single atom losses that can be of several seconds up to a few minutes, setting an upper limit on the length of the quantum circuit~\cite{Mil2019}. 

The collective spins are coupled to the phonons through contact interaction according to the scattering length $a^{AB}_0 = 756 a_0$, $a^{AB}_1 = 2542 a_0$ and $ a^{AB}_2 = -437  a_0$.
The couplings $g^{m}_{AB}$ can then
be obtained by calculating the corresponding Clebsch-Gordon coefficients \cite{Kawaguchi2012}.
The speed of the entangling gates is then determined by the strength of the interaction between species A and B given by Eq.~\eqref{eq:effectiveInteraction}, as we illustrated in Fig.~\ref{fig:SpinPhonoModel1D} and Fig.~\ref{fig:SpinPhonoModel2D}.

As these numbers show, this proposal provides a realistic route  for large-scale quantum information processing in neutral atoms, exploiting already available state-of-the-art technology.

\section{Quantum error correction \label{sec:QEC} }
Quantum error correction (QEC) allows one to mitigate the effects of a noisy environment and faulty operations on the information stored in quantum states. 
The main idea of QEC is to embed quantum information
in a Hilbert space of larger dimension, enabling a distribution of information that leads to resilience against noise. 
For this purpose, one can use the combined Hilbert space of multiple qubits, while an alternative approach is to use a single system that has a larger Hilbert space. 
A way to realize the latter goes back to a seminal work of Gottesmann, Kitaev, and Preskill (GKP)~\cite{GKP2001}, which proposed encoding a qubit into a harmonic oscillator. 

The {\em finite}-dimensional version of the GKP code encodes a qubit into a collective spin. 
This code is based on a set of commuting operators, the {\em stabilizer set},
which can be measured simultaneously. 
The quantum information is then encoded into the joint $(+1)$-eigenspace of the stabilizer set.
Errors acting on the encoded information may lead to a change in one or few stabilizer measurements, 
which helps in detecting and correcting  errors.

The formalism of the finite-dimensional GKP code rests on the generalized Pauli operators $X$ and $Z$ defined by
\begin{subequations}
\begin{align}
    X \ket{j}_d  &= \ket{ (j+1) \text{ mod } d }_d \,, \\
    Z \ket{j}_d   &= \omega^{j} \ket{j}_d \, .
\end{align}
\end{subequations}
These operators obey the relation $Z X=\omega X Z$, with $\omega=\exp (2 \pi i / d)$. 
In particular, two operators of the form $X^\alpha Z^\beta$ and $X^\gamma Z^\delta$ can commute with each other, if their exponents  $\alpha$, $\beta, \gamma$, and $\delta$ 
are chosen appropriately. Such operators can then be used to form the stabilizer group of a code.


To illustrate the idea behind the finite dimensional GKP code, we choose $d~=~8$ corresponding to $7$ atoms in an optical tweezer. 
In order to generate the stabilizer group, we choose the operators
\begin{subequations} 
\begin{align}
    S_1 &=X^4 \,, \\
    S_2 &=Z^4 \,.
\end{align}
\end{subequations}
Notice that $S_1$ and $S_2$ commute. 
Their joint $(+1)$-eigenspace defines a two-dimensional subspace, which can be used to encode a logical qubit. 
A basis for this code space is
\begin{subequations} \label{eq:Codewords}
\begin{align}
    \ket{\overbar{0}} &= \tfrac{1}{\sqrt{2}} (\ket{0}_8 + \ket{4}_8) \, , \label{eq:BasisState0} \\
    \ket{\overbar{1}} &= \tfrac{1}{\sqrt{2}} (\ket{2}_8 + \ket{6}_8 ) \,.
\end{align}
\end{subequations}
\begin{figure}
    \centering
    \includegraphics[width=\columnwidth]{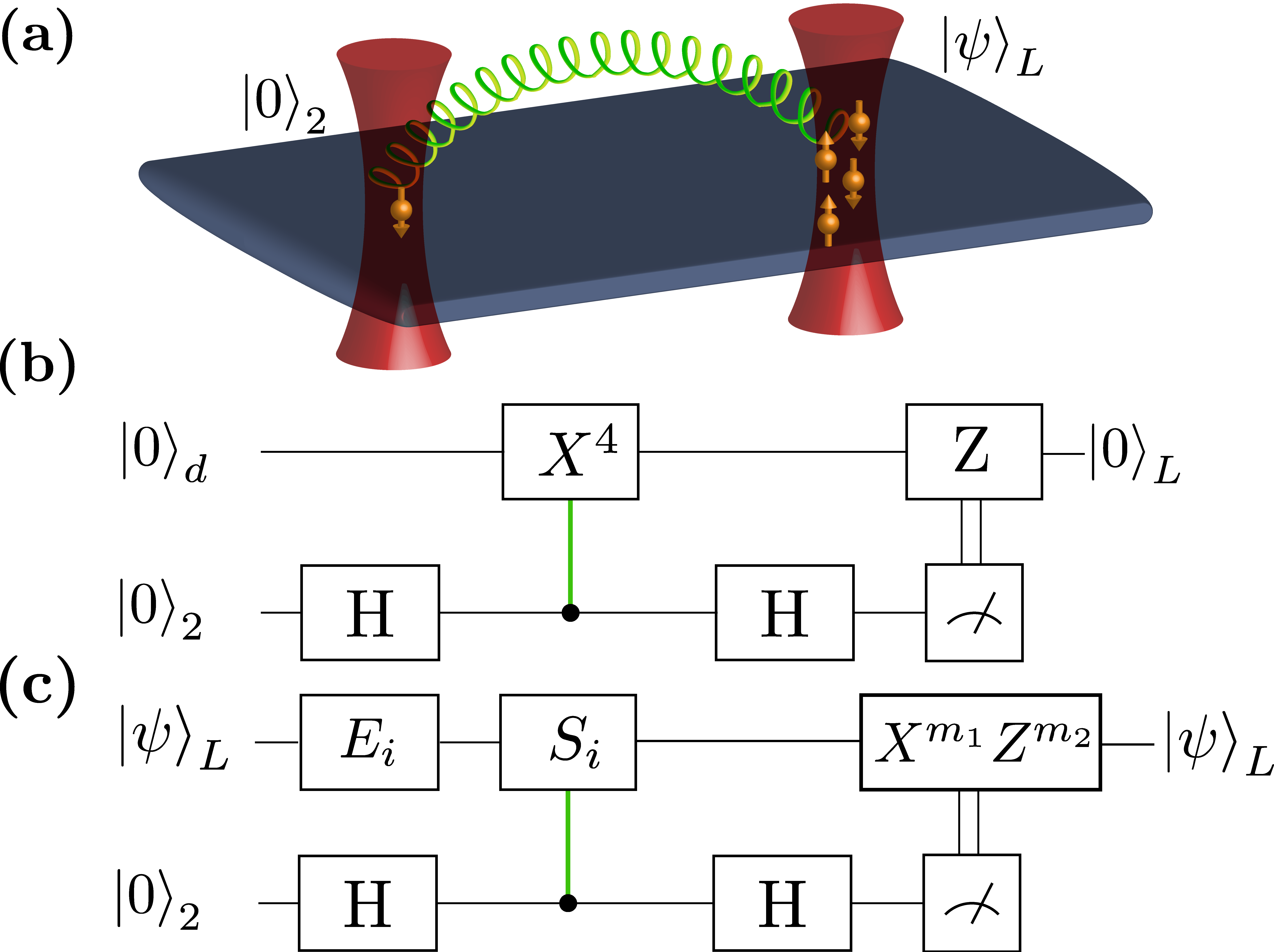}
    \caption{\textbf{Error correction: (a)} Experimental implementation with the tweezer setup. {The phonons of the bosonic bath couple a tweezer with $N_A$ atoms (providing the $d=N_A+1$-dimensional Hilbert space to encode a logical qubit) to a tweezer with one atom (providing a control qubit required for state preparation and error detection).}
    \textbf{(b)} Preparation of the logical state $\ket{\overbar{0}}$
    given in Eq.~\eqref{eq:Codewords} using a control qubit, Hadamard gates, a $\operatorname{CX}^4$ gate {, and a Z rotation conditioned on the outcome of the measurement on the control qubit.}
    \textbf{(c)} Error detection and correction on an encoded state $\ket{\psi}_L$. The GKP code $d = 8$ with $S_1= X^4$ and $S_2 = Z^4$ can detect the errors $Z$ and $X$, and no feedback of the measurement result is needed. However, the feedback becomes essential for the correction of errors e.g. for the GKP code in $d = 8$ with $S_1= X^6$ and $S_2 = Z^6$.}
    \label{fig:PrepAndCorrect}
\end{figure}
%
A state is then encoded as $\ket{\psi_L} = \alpha \ket{\overbar{0}} + \beta \ket{\overbar{1}}$ and one can 
detect all errors $X^aZ^b$ that have $|a|,|b| \leq 1$.\ 
However, from the fact that $X\ket{\bar{0}} = X^{-1} \ket{\bar{1}}$ we infer that some of these errors can only be detected, but not corrected, because the action of $X$ and $X^{-1}$ on the logical states cannot be distinguished. More explicitly, this can be seen from the conditions for quantum error correction~\cite{Knill1997}.

Experimentally, the two logical states $\ket{\bar{0}}, \ket{\bar{1}}$  are prepared
by using a control qubit 
as shown in Fig.~\ref{fig:PrepAndCorrect}b: The collective spin and the qubit are initially 
prepared in the product state $\ket{0}_2 \ket{0}_8$. Applying a Hadamard gate onto the control qubit 
leads to  $\frac{1}{\sqrt{2}}  (\ket{0}_2 + \ket{1}_2)  \ket{0}_8 $. 
Next, we employ a conditional $\operatorname{CX}^4$ gate defined as
\begin{align}
    \operatorname{CX}^4 \ket{j}_2 \ket{k}_8  = (\mathds{1} \otimes X^{4j}) \ket{j}_2 \ket{k}_8 \,,
\end{align}
which results in $\frac{1}{\sqrt{2}} (\ket{0}_2 \ket{0}_8 + \ket{1}_2 \ket{4}_8 )$.
A second Hadamard gate on the control qubit yields
\begin{align}
   \frac{1}{\sqrt{2}} \left[ \ket{0}_2 \ket{\overbar{0}} +   \ket{1}_2 (Z\ket{\overbar{0}}) \right] \,.
\end{align}
After measuring the $Z$ eigenvalue of the control qubit, the collective spins is 
projected onto either $\ket{\overbar{0}}$  or $Z \ket{\overbar{0}}$. 
{In the latter case, an additional $Z$ gate is applied. In this way, the system is deterministically prepared in $\ket{\overbar{0}}$.}

The encoding \eqref{eq:Codewords} allows one to detect certain errors affecting the logical qubit.
Consider errors of the form $X^a Z^b$ acting on $\ket{\psi_L}$. From the commutation relation $S_1 X^a Z^b = \omega^{-4 a} X^a Z^b S_1$ and $S_2 X^a Z^b = \omega^{-4 b} X^a Z^b S_2$, one obtains
\begin{subequations}
\begin{align} \label{Eq:Changeof }
    S_1  \left( X^a Z^b \ket{\psi_L} \right) &=  
     {\omega}^{-4b} \left( X^a Z^b  \ket{\psi_{L}} \right) \, ,\\
    S_2  \left( X^a Z^b \ket{\psi_{L}}\right) &=  
    {\omega}^{-4 a}  \left( X^a Z^b \ket{\psi_{L}} \right) \, . 
\end{align}
\end{subequations}
Thus an error may move the logical qubit out of the $(+1)$-eigenspace of $S_1$ and $S_2$.
In particular, this happens for all non-trivial $X^a Z^b $ with $|a|,|b|\leq 1$.
The circuit in Fig.~\ref{fig:PrepAndCorrect}c can then be used to detect when this happens.
Likewise, it can be shown that the same circuit also detects any linear combinations of such errors.

To also correct errors, one has to use a larger collective spin. Using $d=18$, the stabilizers
$S_1 = X^6$ and $S_2 = Z^6$
yield a code with basis~\cite{GKP2001}
\begin{subequations}
\begin{align}
    \ket{\overbar{0}} &= \tfrac{1}{\sqrt{3}} \left( \ket{0}_{18} + \ket{6}_{18} + \ket{12}_{18} \right) \, , \\
    \ket{\overbar{1}} &= \tfrac{1}{\sqrt{3}} \left( \ket{3}_{18} + \ket{9}_{18} + \ket{15}_{18} \right) \, .
\end{align}
\end{subequations}
The circuit given in Fig.~\ref{fig:PrepAndCorrect}d can now not only detect, but also {\em correct} all errors $X^a Z^b$  with $|a|,|b|\leq 1$.

Then, the encoded states can also be manipulated through a set of universal gates. 
To be specific, we return to the case of $d=8$. The logical gates $\overbar{X} = X^2$ and $\overbar{Z} = Z^2$ act like the usual spin $1/2$ Pauli matrices on $\ket{\overbar{0}}$ and $\ket{\overbar{1}}$, i.e.,
\begin{subequations}
\begin{align}
    \bar{X} \ket{\overbar{0}} &= \ket{\overbar{1}}\,, &
    \bar{Z} \ket{\overbar{0}} &=  \ket{\overbar{0}} \,, \\
    \bar{X} \ket{\overbar{1}} &= \ket{\overbar{0}}\,, &
    \bar{Z} \ket{\overbar{1}} &= -\ket{\overbar{1}}\,,
\end{align}
\end{subequations}
such that $\bar{X}^2 = \bar{Z}^2 = \mathbb{1}$ and $\bar{X} \bar{Z} + \bar{Z}\bar{X} = 0$.

As explained in Sec.~\ref{sec:EffectiveSpins}, 
we can implement 
{\em all} single-qubit logical gates
through Trotterization, which allows to synthesize
\begin{align}
    \bar{\mathcal{R}}(\alpha, \beta, \gamma, \delta)      &= e^{i\alpha}e^{i \beta \bar{Z}} e^{i \gamma \bar{X}} e^{i \delta \bar{Z}}.
\end{align}
Moreover, two logical qubits can be entangled by using a logical controlled Z gate,
\begin{align}
    \operatorname{C\bar{Z}}(\mathbf{x}, \mathbf{y}) &= e^{i \pi \bar{Z}(\mathbf{x}) \bar{Z}(\mathbf{y})} \,,
\end{align}
which can be also synthesized via Trotterzation using the gates generated by $L_x$, $L_z$ and $L^2_z$, and the entangling gate $\operatorname{CZ}(\mathbf{x},\mathbf{y})$ given in Eq.\eqref{eq:CZgate}.
{Together}, the operations $\bar{\mathcal{R}}(\alpha, \beta, \gamma, \delta)$ and $\operatorname{C\bar{Z}}$ 
then 
generate a universal set of logical gates~\cite{LLoyd1995}. 

In summary, 
we have constructed an explicit example of a
quantum error-detecting code for $d=8$, described the experimental preparation of its logical states, and provided a universal gate set to manipulate the stored quantum information. 
The presented formalism will also work in larger dimensions, 
allowing for the detection 
and correction of a larger set of errors~\cite{GKP2001}.
The here proposed platform with a suitable quantum error-correcting 
scheme, such as finite GKP codes, may allow one to go beyond the abilities of noisy intermediate-scale quantum (NISQ) technologies.   

\section{Conclusion \& Outlook}
In this work, we have proposed a mixture of two ultracold atomic species as a platform for universal quantum computation. Our proposed implementation uses long-range entangling gates and, in addition, allows for the realization of quantum error correction. The presented spin-phonon system is not only useful for the processing of quantum information but is also interesting from a quantum many-body perspective. The many body systems realized by our platform is similar to gauge theories coupled to a Higgs field~\cite{Fradkin1979, Gonzalez-Cuadra2017, VanDamme2020} and hence is a promising candidate for the investigation of topological matter. Further, the here proposed spin-phonon systems may give access to the physics of the Peierls transition \cite{ z2_bhm_3}, the Jahn-Teller effect \cite{Porras2012} in a many-body context as well as the study of frustrated spin models \cite{Nevado2016}. The versatility of this platform also makes it an ideal candidate for a fully programmable quantum simulation, whose large potential has been demonstrated in Rydberg and trapped ion systems~\cite{Bernien2017, Zhang2017,Kokail2019}. For example, atomic mixtures have been proposed for the quantum simulation of quantum chemistry problems~\cite{Luengo2020}.

In future work, the building blocks of our proposed platform could be exchanged ad libitum exploiting the versatility of atomic, molecular, and atomic physics. The basic unit of information can be stored, e.g., with spinful fermions in an optical tweezer~\cite{Bergschneider2019} or higher dimensional spins~\cite{ Aikawa2012, Stamper-Kurn2013, Chalopin2020}. The entanglement bus may be substituted by more complex many-body excitations like magnons or rotons~\cite{Fukuhara2013, Chomaz2018}. This exchange of the basic unit of information and the entanglement mechanism has the  potential to lead to new implementations of quantum algorithms and quantum error correction schemes, see for example~\cite{Grimsmo2020,Noh2020,Albert2020, Gross2020}. Thus, ultracold atom mixtures present a promising platform to implement fault-tolerant quantum computation in a scalable platform, paving a way beyond the abilities of noisy intermediate-scale quantum technology.

\section{Acknowledgments}
The authors are grateful for fruitful discussions with J. Eisert, M. Gaerttner, T. Gasenzer, M. Gluza, S. Jochim, M. Oberthaler, A.P. Orioli, P. Preiss, H. Strobel, E. Tiemann and  all the members of the SynQs seminar.
ICFO group acknowledges support from ERC AdG NOQIA, Spanish Ministry of Economy and Competitiveness (“Severo Ochoa” program for Centres of Excellence in R\&D (CEX2019-000910-S), Plan National FISICATEAMO and FIDEUA PID2019-106901GB-I00/10.13039 / 501100011033, FPI), Fundació Privada Cellex, Fundació Mir-Puig, and from Generalitat de Catalunya (AGAUR Grant No. 2017 SGR 1341, CERCA program, QuantumCAT \textunderscore U16-011424, co-funded by ERDF Operational Program of Catalonia 2014-2020), MINECO-EU QUANTERA MAQS (funded by State Research Agency (AEI) PCI2019-111828-2 / 10.13039/501100011033), EU Horizon 2020 FET-OPEN OPTOLogic (Grant No 899794), and the National Science Centre, Poland-Symfonia Grant No. 2016/20/W/ST4/00314.  This project has received funding from the European Union's Horizon 2020 research and innovation programme under the Marie Skłodowska-Curie grant agreement No. 754510. 

A.D. acknowledges financial support from a fellowship granted by la Caixa Foundation (ID 100010434, fellow-ship code LCF/BQ/PR20/11770012).
  F.H. acknowledges 
  the Government of Spain (FIS2020-TRANQI and Severo Ochoa CEX2019-000910-S), 
  Fundació Cellex, 
  Fundació Mir-Puig, 
  Generalitat de Catalunya (CERCA, AGAUR SGR 1381).
  
P.H. acknowledges support by Q@TN --- Quantum Science and Technologies at Trento, the Provincia Autonoma di Trento, and the ERC Starting Grant StrEnQTh (Project-ID 804305). 

This work is part of and supported by the DFG Collaborative Research Centre “SFB 1225 (ISOQUANT)”. F. J. acknowledges the DFG support through the project FOR 2724, the Emmy- Noether grant (Project-ID 377616843) and support by the Bundesministerium für Wirtschaft und Energie through the project "EnerQuant" (Project-ID 03EI1025C).

\appendix

\section{Collective spin Hamiltonian \label{sec:SingleHamiltonian}}
The type A atoms are placed in an external magnetic field, which allows to prepare the atoms in magnetic substates $m \in \{ 0,1 \}$. Further, the atoms of type A are localized at specific sites via 
optical potentials (see Fig.~\ref{fig:Overview}). The single particle physics is described by the Hamiltonian 
\begin{align}\label{Eq:HamSinglePart}
  H^{m}_{A}(\mathbf{x})=-\frac{{{\hbar}^{2}}\nabla _{\mathbf{x}}^{2}}{2{{M}_{A}}}+V_A(\mathbf{x})+E^{m}_{A}(B)  \, ,
\end{align}
where  ${M}_{A}$ is the mass of 
the atoms and  ${{V}_{A}}\left( \mathbf{x} \right)$ is the optical potential confining the atoms. The energy shift $E_{m}^{A}(B)$ determined by the magnetic field B is given by 
\begin{align}
  E^{A}_{m}(B)=p_{m}(B)m+q_{m}(B){{m}^{2}} \, , 
\end{align}
where $p_m(B)$ parametrizes the linear Zeeman shift and $q_m(B)$ the quadratic Zeeman shift. 
The field operators ${A}_{m}(\mathbf{x})$ and $A_{m}^{\dagger}(\mathbf{x})$ annihilate and create a particle at position $\mathbf{x} = (x_1, \ldots, x_d )^T$ respectively and fulfill canonical commutation relations. The many-body Hamiltonian is given by
\begin{align}
    H_{A}&= \sum_{m} \int_{\mathbf{x}} A_{m}^{\dagger}(\mathbf{x}) H^{m}_{A}(\mathbf{x}) A_{m}(\mathbf{x}) \notag \\
	 &+\sum_{m,n} \int_{\mathbf{x}} \frac{g^{mn}_{A}}{2}A_{m}^{\dagger } (\mathbf{x})A_{n}^{\dagger }(\mathbf{x}){A}_{n} (\mathbf{x}) {A}_{m} (\mathbf{x}) \, .
\end{align}
The atoms are localized at different sites $\mathbf{y}$, which allows one to 
expand the field operators as
\begin{align} \label{eq:ExpansionOfA}
    A_{m}(\mathbf{x}) = \sum_{\mathbf{y}} \varphi_A(\mathbf{x}-\mathbf{y}) a_m ( \mathbf{y}) \, 
\end{align}
with a wave function $\varphi_A$ localized at $\boldsymbol{0}$ and 
does not depend on the magnetic substate. 
Inserting the expansion \eqref{eq:ExpansionOfA} into $H_{A}$ 
and neglecting the tunneling terms, we obtain the Hamiltonian 
\begin{align}
    H_A &=  \frac{1}{2}\sum_{m,n,\mathbf{y}}  \tilde{g}_A^{mn} {a}^{\dagger}_{m}(\mathbf{y}){a}^{\dagger}_{n}(\mathbf{y}){a}^{\vphantom{\dagger}}_{m}(\mathbf{y}){a}_n^{\vphantom{\dagger}}(\mathbf{y}) \,  \label{eq:H_A_reduced}
\end{align}
with the overlap integrals
\begin{align}
\tilde{g}_A^{mn} = g_A^{mn} \int_\mathbf{x} |\varphi_A (\mathbf{x} - \mathbf{y})|^4  \,.
\end{align}
To be explicit, we approximate $\varphi_A (\mathbf{x})$
by the ground state wavefunction of an isotropic harmonic oscillator $\varphi_A(\mathbf{x}) = 
\varphi_A (x_1) \varphi_A (x_2) \varphi_A (x_3)$ with
\begin{align}
    \varphi_A (x_i)  =   (\sqrt{\pi} \sigma_A)^{-1/2} e^{-\frac{1}{2} \left( \frac{x_i}{\sigma_A}\right)^2} \label{eq:Harmonic_Oscillator_WF}
\end{align}
with the characteristic length $\sigma_A$. 
Calculating the overlap integral leads to the dimensional reduced coupling constant
\begin{align}
\tilde{g}_A^{mn} = {(2\pi)^{-3/2}} (g_A^{mn}/\sigma_A^3)  \, .
\end{align}
Inserting the Schwinger representation of the angular momentum into Eq.~\eqref{eq:H_A_reduced}
leads to Eq.~\eqref{eq:HamiltonianA} with the coupling constants
\begin{subequations}
\begin{align}
    \chi &= \frac{1}{2} \left( \tilde{g}_A^{00} + \tilde{g}_A^{11}  -2 \tilde{g}_A^{10}  \right) \, , \\
    \Delta &=  \frac{1}{2}(N_A-1) (\tilde{g}_A^{11} - \tilde{g}_A^{00}) + E^1_A - E^0_A \, ,
\end{align}
\end{subequations}
where $N_A$ is the particle number on each site.
\section{Enveloping algebra of U(3) \label{sec:EnvelopingAlgebra}}
In this appendix, we demonstrate for $\ell =1$  that $L_z$, $L_x$, and $L^2_z$ are sufficient  
to generate all $U(3)$ matrices. 
We use the spin matrices 
\begin{align}
    L_x = \frac{1}{\sqrt{2}} 
    \begin{psmallmatrix}
        0 & 1 & 0 \\ 
        1 & 0 & 1 \\
        0 & 1 & 0
    \end{psmallmatrix}, 
    L_y =  \frac{1}{\sqrt{2}i} 
    \begin{psmallmatrix}
        0  & 1 & 0 \\ 
        -1 & 0 & 1 \\
         0 & -1 & 0
    \end{psmallmatrix}, \notag 
    L_z =   
    \begin{psmallmatrix}
        1  & 0 & 0 \\ 
        0 & 0 & 0 \\
        0 & 0 & -1
    \end{psmallmatrix}
\end{align}
Consider the following matrices generated by commutators of $L_x$, $L_z$, and $(L_z)^2$
\begin{align}
    \mathcal{M}_1 &= L_x,         & \mathcal{M}_2 & = L_z,          & \mathcal{M}_3 &= L_z^2,        \notag \\
    \mathcal{M}_4 &= i[\mathcal{M}_1,\mathcal{M}_2] ,       & \mathcal{M}_5 &= i[\mathcal{M}_3, \mathcal{M}_1],   & \mathcal{M}_6 &= i[\mathcal{M}_3,\mathcal{M}_4], \notag\\
    \mathcal{M}_7 &= i[\mathcal{M}_5,\mathcal{M}_1],  & \mathcal{M}_8 &= i[\mathcal{M}_5, \mathcal{M}_4],   & \mathcal{M}_9 &= i[\mathcal{M}_6, \mathcal{M}_4] \, .
\end{align}
These commutators form a basis for the Lie algebra of U(3). This can be explicitly checked by
constructing a change of basis from the $\{\mathcal{M}_i\}_{i=1}^9$ to the canonical basis of 
Hermitian matrices $\{M_i\}_{i=1}^9$ given by
\begin{align}
    M_1 &= \begin{psmallmatrix}
        1 & 0 & 0 \\
        0 & 0 & 0 \\
        0 & 0 & 0
    \end{psmallmatrix}, &
    M_2 &= \begin{psmallmatrix}
        0 & 0 & 0 \\
        0 & 1 & 0 \\
        0 & 0 & 0
    \end{psmallmatrix}, & 
    M_3 &= \begin{psmallmatrix}
        0 & 0 & 0 \\
        0 & 0 & 0 \\
        0 & 0 & 1
    \end{psmallmatrix}, &  \notag \\
    M_4 &= \begin{psmallmatrix}
        0 & 1 & 0 \\
        1 & 0 & 0 \\
        0 & 0 & 0
    \end{psmallmatrix}, &
    M_5 &= \begin{psmallmatrix}
        0 & 0 & 1 \\
        0 & 0 & 0 \\
        1 & 0 & 0
    \end{psmallmatrix}, & 
    M_6 &= \begin{psmallmatrix}
        0 & 0 & 0 \\
        0 & 0 & 1 \\
        0 & 1 & 0
    \end{psmallmatrix}, & \notag \\
    M_7 &= \begin{psmallmatrix}
        0 & i & 0 \\
        -i & 0 & 0 \\
        0 & 0 & 0
    \end{psmallmatrix}, &
    M_8 &= \begin{psmallmatrix}
        0 & 0 & i \\
        0 & 0 & 0 \\
        -i & 0 & 0
    \end{psmallmatrix}, & 
    M_9 &= \begin{psmallmatrix}
        0 & 0 & 0 \\
        0 & 0 & i \\
        0 & -i & 0
    \end{psmallmatrix}.
\end{align}
\section{Phonon Hamiltonian\label{Sec:Phonon}}
The Hamiltonian involving  only the atomic species B is given by
\begin{align}
    H_{B}= &\int_{\mathbf{x}} B^{\dagger}(\mathbf{x}) H^{0}_{B}(\mathbf{x}) B(\mathbf{x})  +\frac{g_{B}}{2}  \int_{\mathbf{x}} B^{\dagger } (\mathbf{x})B^{\dagger }(\mathbf{x}){B} (\mathbf{x}) {B} (\mathbf{x})  \, 
\end{align}
with 
\begin{align}\label{Eq:HamSinglePart}
  H^{0}_{B}(\mathbf{x})=-\frac{{{\hbar}^{2}}\nabla _{\mathbf{x}}^{2}}{2{{M}_{B}}}+V_B(\mathbf{x})  \, .
\end{align}
In order to create a one- or two-dimensional phononic bath we apply the following harmonic and isotropic confinement 
\begin{align}
  V_B(\mathbf{x}) & = \frac{1}{2}M_B \omega^2_B \sum_{i=n+1}^d  x_i^2  \, 
\end{align}
For sufficiently low temperatures, the transversal degress of freedom will not be excited, which will confine the particles effectively to one ($n=1$) or two ($n=2$) dimensions  respectively. This freezing out 
of the transversal directions allows one to write the field operator as 
\begin{align}
    B(\mathbf{x})  &=  B(x_{1},...,x_n) \varphi_{B}(x_{n+1},...,x_d)  \label{eq:dimreduced}
\end{align}
with the transversal wave function $\varphi_{B}(x_{n+1},...,x_d)$ and
$B(x_{1},...,x_n) $ the annihiliation operator in $n$ dimensions, and
in the following we will use $\tilde{\mathbf{x}} = (x_1, \ldots, x_n)^T$.
The stationary Gross-Pitaevskii equation is given by
\begin{align}
    0 = \left[ -\frac{\hbar^2\nabla_{\tilde{\mathbf{x}}}^2}{2M_B} - \mu_B + \tilde{g}_B |\phi_B(\tilde{\mathbf{x}})|^2 \right] \phi_B(\tilde{\mathbf{x}})
\end{align}
with the condensate $\phi_B(\tilde{\mathbf{x}})$ fulfilling the Dirichlet boundary conditions $\phi_B(\tilde{\mathbf{x}}) = 0$  for
$\tilde{\mathbf{x}} \in \partial D$ with $D = [0, L]^n$ and chemical potential $\mu_B$. The coupling constant is given by
\begin{align}
    \tilde{g}_B = g_B \int_{x_{n+1}, \ldots, x_d}  |\varphi_{B}(x_{n+1},...,x_d)|^4 \, ,
\end{align}
which becomes $\tilde{g}_B = (\sqrt{2\pi}\sigma_B)^{-(d-n)} g_B$ for harmonic confinement.
The bulk solution of the Gross-Pitaevskii equation can be approximated by the homogeneous function
\begin{align}
    \phi_B(\tilde{\mathbf{x}}) &\approx \sqrt{\frac{\mu_B}{\tilde{g}_B}} \,
\end{align}
leading to the density $n_B  = \frac{\mu_B}{\tilde{g}_B}$.
In order to study the excitations of the bulk solution, we perform the Bogoliubov approximation 
\begin{align}
    B(\tilde{\mathbf{x}})     &= \phi_B(\tilde{\mathbf{x}})  + \delta B(\tilde{\mathbf{x}}) \, , 
    \label{eq:DefFluctuations}
\end{align}
where $\delta B$ and $\delta B^{\dagger}$ fulfill canonical commutation
relations. 
The Bogoliubov Hamiltonian approximation of $H_B$ is given by
\begin{align}
H_B = \int_{D} &\left[ \frac{\hbar^2}{2M_{\text{B}}} | \nabla_{\tilde{\mathbf{x}}} \delta B|^2 - \mu_B  |\delta B|^2  + 2\tilde{g}_B |\delta B|^2  |\phi_{B}|^2  \right. \notag \\
&\left. + \frac{\tilde{g}_B}{2} \left[  (\delta B^{\dagger})^2 (\phi_{B})^2  +  (\delta B)^2 (\phi^{\ast}_{B})^2 \right] \right] \, .
\end{align}
The equations of motions can be solved by using the mode expansion
\begin{align}\label{eq:expansion}
\delta B(\tilde{\mathbf{x}},t) = \sum_{\mathbf{k}} \left[ {b}_{\mathbf{k}} u_{\mathbf{k}}(\tilde{\mathbf{x}}) e^{-i\omega_{\mathbf{k}} t} + {b}^{\dagger}_{\mathbf{k}} v^{\ast}_{\mathbf{k}}(\tilde{\mathbf{x}}) e^{i\omega_{\mathbf{k}}t} \right] \, , 
\end{align}
where the sum does not include the condensate mode~\cite{Rogel_Salazar_2001}.
This expansion leads to the following generalized eigenvalue problem
\begin{align}
    \omega_{\mathbf{k}}\begin{pmatrix}
    1 & 0 \\
    0 & -1
    \end{pmatrix}
    \begin{pmatrix}
    u_\mathbf{k} \\
    v_\mathbf{k}
    \end{pmatrix}
    = 
    \begin{pmatrix}
    h(\tilde{\mathbf{x}})                               & \tilde{g}_B  (\phi_B)^2              \\
\tilde{g}_B (\phi_B^{\ast})^2    & h(\tilde{\mathbf{x}})
    \end{pmatrix}
    \begin{pmatrix}
    u_\mathbf{k} \\
    v_\mathbf{k}
    \end{pmatrix} \, ,
\end{align}
with
\begin{align}
    h(\tilde{\mathbf{x}}) =  -\frac{\hbar^2\nabla_{\tilde{\mathbf{x}}}^2}{2M_{B}}  - \mu_B  + 2\tilde{g}_{B} |\phi_{B}(\tilde{\mathbf{x}})|^2 \,.
\end{align}
Solving this generalized eigenvalue problem will lead to the orthonormalization condition
\begin{align}
    \int_D \left[ u^{\ast}_{\mathbf{k}}(\tilde{\mathbf{x}}) u_{\mathbf{k}'}(\tilde{\mathbf{x}}) -  v^{\ast}_{\mathbf{k}}(\tilde{\mathbf{x}}) v_{\mathbf{k}'}(\tilde{\mathbf{x}}) \right] = \delta_{\mathbf{k},\mathbf{k}'} \,.
\end{align}
Since the background-field is approximately constant and 
because of the Dirichlet boundary conditions, we make the ansatz
\begin{subequations} \label{eq:ModeFunctions}
\begin{align} 
u_{\mathbf{k}}(\tilde{\mathbf{x}}) &= u_{\mathbf{k}} \left(\frac{2}{L}\right)^{n/2} \prod_{i=1}^{n} \sin(k_i x_i)  \, , \\
v_{\mathbf{k}}(\tilde{\mathbf{x}}) &= v_{\mathbf{k}} \left(\frac{2}{L}\right)^{n/2} \prod_{i=1}^{n} \sin(k_i x_i)   \, ,
\end{align} 
\end{subequations}
with $k_i = \frac{n_i \pi}{L}$ and $n_i\geq 2$ and the amplitudes are given by
\begin{subequations}
\begin{align}
    u^2_{\mathbf{k}}  &= \frac{1}{2}\left({\frac{\varepsilon_{\mathbf{k}}}{2\omega_{\mathbf{k}}} + 1} \right)  \, , \\ 
    v^2_{\mathbf{k}}  &= \frac{1}{2}\left({\frac{\varepsilon_{\mathbf{k}}}{2\omega_{\mathbf{k}}} - 1} \right)  \, ,
\end{align}
\end{subequations}
with $\varepsilon_{\mathbf{k}} = (\hbar^2 \mathbf{k}^2) /(2M_B) + \tilde{g}_B |\phi_B|^2 $.
The Bogoliubov eigenfrequencies of the excitations are given by 
\begin{align}
    \hbar \omega_{\mathbf{k}} =  \sqrt{ \frac{\hbar^2{\mathbf{k}}^2}{2M_B} 
    \left(\frac{\hbar^2\mathbf{k}^2}{2 M_B}  +2 \mu_B\right)} \, .
\end{align}

\section{Spin-phonon interaction \label{app:SpinPhonon}}
In order to derive the interaction between the phonons and the collective spins,
we start from Hamiltonian modeling the interaction between the A and B atoms 
\begin{align}
    H_{AB} = \sum_m \int_{\mathbf{x}} \frac{g^m_{AB}}{2} A_m^{\dagger }(\mathbf{x})A_m(\mathbf{x})B^{\dagger}(\mathbf{x}) B (\mathbf{x}) \,.
\end{align}
 After reducing the dimension and considering the tight confinement of the spins, i.e.,  Eq.~\eqref{eq:dimreduced} and Eq.~\eqref{eq:ExpansionOfA}, we obtain the Hamiltonian
\begin{align}
    &H_{AB} =  \sum_{m,\mathbf{y}} \frac{\tilde{g}^m_{AB}}{2} a_m^{\dagger }(\mathbf{y})a_m(\mathbf{y}) \notag \\
    &\times \int_{D} |\varphi_A(x_{1} - y_{1}) \ldots \varphi_A(x_{n} - y_{n})|^2B^{\dagger}(\tilde{\mathbf{x}}) B (\tilde{\mathbf{x}}) \, ,
\end{align}
where we neglected hopping of the A atoms and introduced the coupling constant
\begin{align}
    \tilde{g}^m_{AB} &= {g}^m_{AB} \int_{x_{n+1}, \ldots, x_d} |\varphi_{B}(x_{n+1})...\varphi_{B}(x_d)|^2 \notag \\
&\times |\varphi_A(x_{n+1} - y_{n+1}) \ldots \varphi_A(x_{d} - y_{d})|^2 \, ,
\end{align}
which can be written as
\begin{align}
    \tilde{g}^m_{AB} &=  {g}^m_{AB} \left[ \pi (l^2_A + l^2_B) \right]^{-(d-n)/2}
\end{align}
for  harmonic confinement of the A and B atoms with harmonic oscillator length scale $\sigma_A$ and
$\sigma_B$ respectively. 
Expanding the field operator B  in fluctuations as in Eq.~\eqref{eq:DefFluctuations} and neglecting terms of order $\mathcal{O}(\delta B^2)$ one obtains a Hamiltonian
\begin{align}
    H_{AB} = H^{(0)}_{AB} + H^{(1)}_{AB} \, ,
\end{align}
where $ H^{(0)}_{AB}$ is independent of the fluctuations 
and $ H^{(1)}_{AB}$ is linear in the fluctuations.
The first contribution is given by
\begin{align}
     H^{(0)}_{AB} =  \sum_{m,\mathbf{y}} \Delta_m a_m^{\dagger }(\mathbf{y})a_m(\mathbf{y})  \, ,
\end{align}
with the coupling constant 
\begin{align}
    \Delta_m =&  \frac{1}{2}\tilde{g}^m_{AB} n_B \int_{D} |\varphi_A(x_{1} - y_{1}) \ldots \varphi_A(x_{n} - y_{n})|^2  \, ,
\end{align}
and since the harmonic oscillator wave functions are normalized we obtain
\begin{align}
    \Delta_m =&  \frac{1}{2}\tilde{g}^m_{AB} n_B  \, .
\end{align}
The contribution linear in the fluctuations is given by 
\begin{align}
    H_{AB} &=  \frac{\sqrt{n_B}}{2} \sum_{m,\mathbf{y}}  \tilde{g}^m_{AB} a_m^{\dagger }(\mathbf{y})a_m(\mathbf{y}) \int_{D}  [\delta B(\tilde{\mathbf{x}})  +\text{H.c.}] \notag \\
    &\times  |\varphi_A(x_{1} - y_{1}) \ldots \varphi_A(x_{n} - y_{n})|^2 \, .
\end{align}
Inserting the mode expansion \eqref{eq:expansion} into $H^{(1)}_{AB}$ we obtain
\begin{align}
    H^{(1)}_{AB} &=  \sum_{m,\mathbf{y}, \mathbf{k}} \tilde{g}^m_{AB, \mathbf{k}}(\mathbf{y}) a_m^{\dagger }(\mathbf{y})a_m(\mathbf{y})  \left[ b_{\mathbf{k}} + \text{H.c.} \right]   \, ,
\end{align}
where we introduced the coupling constant
\begin{align}
\tilde{g}^m_{AB, \mathbf{k}}(\mathbf{y}) &= \frac{\sqrt{n_B}}{2}   \tilde{g}^m_{AB} \int_{D}  |\varphi_A(x_{1} - y_{1}) \ldots \varphi_A(x_{n} - y_{n})|^2 \notag \\
&\times [u_{\mathbf{k}}(\tilde{\mathbf{x}}) + v_{\mathbf{k}}(\tilde{\mathbf{x}})] \, . \label{eq:SpinPhononCoupling1}
\end{align}
Inserting \eqref{eq:ModeFunctions} and \eqref{eq:Harmonic_Oscillator_WF} we obtain for
$L\gg \sigma_B$ approximately
\begin{align}
\tilde{g}^m_{AB, \mathbf{k}}(\mathbf{y}) &=    \tilde{g}^m_{AB} \sqrt{n_B}  \left( \frac{2}{L}\right)^{n/2}
(u_{\mathbf{k}} + v_{\mathbf{k}}) e^{-\frac{1}{4} (\mathbf{k} l_{\!\scriptscriptstyle{A}})^2} \notag \\
&\times  \prod_{i=1}^n \sin(k_i y_i) \, .
\end{align}
Using the Schwinger representation (see Eq.~\eqref{eq:SchwingerRepresentation}), we obtain the interaction between the spins and the phonons 
\begin{align}
    H^{(1)}_{AB} &= \sum_{m,\mathbf{y}, \mathbf{k}} \tilde{g}^m_{AB, \mathbf{k}}(\mathbf{y}) \left[ L(\mathbf{y}) - (-1)^m L_z(\mathbf{y}) \right]  \left( b_{\mathbf{k}} + \text{H.c.} \right) \, ,
\end{align}
with $L(\mathbf{y})$ being the length of the angular momentum on site $\mathbf{y}$. Reshuffling terms leads to Eq.~\eqref{eq:SpinPhononHamiltonian} with 
the coupling constants
\begin{subequations}
\begin{align}
    \bar{g}_{\mathbf{k}}(\mathbf{y}) &= L(\mathbf{y}) [\tilde{g}^0_{AB, \mathbf{k}}(\mathbf{y}) + \tilde{g}^1_{AB, \mathbf{k}}(\mathbf{y})] \, , \\
    \delta g_\mathbf{k}(\mathbf{y})  &=  \tilde{g}^1_{AB, \mathbf{k}}(\mathbf{y}) - \tilde{g}^0_{AB, \mathbf{k}}(\mathbf{y}) \, .
\end{align}
\end{subequations}

\section{Eliminating phonons \label{Sec:AppE}}
Assuming $\Omega(\mathbf{y}) = 0$ and given the approximations of App.~\ref{sec:SingleHamiltonian}, \ref{Sec:Phonon} and \ref{app:SpinPhonon} the Hamiltonian 
$H=H_A + H_B + H_{AB}$ is diagonal in $L_z$, which 
allows us to treat phonons and spins separately. The Heisenberg equation of motion for the phonons is 
\begin{align}
     i \hbar \partial_t b_\mathbf{k} &= \hbar \omega_\mathbf{k} b_\mathbf{k} + \delta b_{\mathbf{k}}  \, 
\end{align}
of $b_{\mathbf{k}}$ operators, where we introduced the abbreviation
\begin{align}
    \delta b_{\mathbf{k}}(\mathbf{y}) =  \sum_{\mathbf{y}} \left[\bar{g}_{\mathbf{k}}(\mathbf{y}) L(\mathbf{y})+  \delta g_\mathbf{k}(\mathbf{y}) L_z(\mathbf{y}) \right]  \, .
\end{align}
We define a shifted annihilation operator as
\begin{align}
    \beta_\mathbf{k} &= b_\mathbf{k} + (\hbar\omega_\mathbf{k})^{-1}\delta b_{\mathbf{k}}  \,.
\end{align}
Inserting the shifted operator in the Hamiltonian Eq.~\eqref{eq:SpinPhononHamiltonian} leads to a spin-spin interaction
\begin{align} \label{Eq:ZZInteractionApp}
H_I =&-\sum_{\mathbf{x}, \mathbf{y}} {g}(\mathbf{x},\mathbf{y}) L_z(\mathbf{x}) L_z(\mathbf{y}) \,,
\end{align}
with
\begin{subequations}
\begin{align}
    g(\mathbf{x},\mathbf{y}) &= \sum_{\mathbf{k}} (\hbar \omega_\mathbf{k})^{-1} \delta g_{\mathbf{k}}(\mathbf{x}) \delta g_{\mathbf{k}}(\mathbf{y}) \, .
\end{align}
\end{subequations}
Inserting the explicit expression for $\delta g_{\mathbf{k}}(\mathbf{x})$, we obtain
Eq.~\eqref{eq:effectiveInteraction}.
\newpage
\bibliographystyle{apsrev4-1}
\bibliography{bibliography}
\end{document}